\documentclass{emulateapj}
\usepackage{lineno}
\usepackage{graphicx}
\usepackage{subfigure}
\usepackage[normalem]{ulem}
\usepackage{amsmath}
\usepackage{upgreek}
\usepackage{hyperref}
\usepackage{breakurl}

\def\arcsec{\hbox{$^{\prime\prime}$}}
\def\arcmin{\hbox{$^{\prime}$}}
\def\deg{\hbox{$^\circ$}}

\def\Fermi{\textit{Fermi}}

\slugcomment{ApJ, accepted}

\shorttitle{{\it Fermi}-LAT Observations of Fornax~A}
\shortauthors{{\it Fermi}-LAT Collaboration}

\begin{document}

\title{{\it Fermi} Large Area Telescope Detection of extended $\gamma$-ray Emission from the Radio Galaxy Fornax~A}

\author{
	M.~Ackermann\altaffilmark{2}, 
	M.~Ajello\altaffilmark{3}, 
	L.~Baldini\altaffilmark{4,5}, 
	J.~Ballet\altaffilmark{6}, 
	G.~Barbiellini\altaffilmark{7,8}, 
	D.~Bastieri\altaffilmark{9,10}, 
	R.~Bellazzini\altaffilmark{11}, 
	E.~Bissaldi\altaffilmark{12}, 
	R.~D.~Blandford\altaffilmark{5}, 
	E.~D.~Bloom\altaffilmark{5}
% ...
}
% ...
\altaffiltext{2}{Deutsches Elektronen Synchrotron DESY, D-15738 Zeuthen, Germany}
\altaffiltext{3}{Department of Physics and Astronomy, Clemson University, Kinard Lab of Physics, Clemson, SC 29634-0978, USA}
\altaffiltext{4}{Universit\`a di Pisa and Istituto Nazionale di Fisica Nucleare, Sezione di Pisa I-56127 Pisa, Italy}
\altaffiltext{5}{W. W. Hansen Experimental Physics Laboratory, Kavli Institute for Particle Astrophysics and Cosmology, Department of Physics and SLAC National Accelerator Laboratory, Stanford University, Stanford, CA 94305, USA}
\altaffiltext{6}{Laboratoire AIM, CEA-IRFU/CNRS/Universit\'e Paris Diderot, Service d'Astrophysique, CEA Saclay, F-91191 Gif sur Yvette, France}
\altaffiltext{7}{Istituto Nazionale di Fisica Nucleare, Sezione di Trieste, I-34127 Trieste, Italy}
\altaffiltext{8}{Dipartimento di Fisica, Universit\`a di Trieste, I-34127 Trieste, Italy}
\altaffiltext{9}{Istituto Nazionale di Fisica Nucleare, Sezione di Padova, I-35131 Padova, Italy}
\altaffiltext{10}{Dipartimento di Fisica e Astronomia ``G. Galilei'', Universit\`a di Padova, I-35131 Padova, Italy}
\altaffiltext{11}{Istituto Nazionale di Fisica Nucleare, Sezione di Pisa, I-56127 Pisa, Italy}
\altaffiltext{12}{Istituto Nazionale di Fisica Nucleare, Sezione di Bari, I-70126 Bari, Italy}
% ...

\begin{abstract}

We report the \Fermi \ Large Area Telescope detection of
extended $\gamma$-ray emission from the lobes of the radio
galaxy Fornax~A using $6.1$ years of Pass~8
data. After Centaurus~A, this is now the second example
of an extended $\gamma$-ray source attributed to a radio galaxy.
Both an extended flat disk morphology and a morphology following the
extended radio lobes were preferred over a point-source description,
and the core contribution was constrained to be $<14$\% of the total $\gamma$-ray flux.
A preferred alignment of the $\gamma$-ray elongation with the
radio lobes was demonstrated by rotating the radio lobes template.
We found no significant evidence for variability on $\sim0.5$ year timescales.
Taken together, these results strongly suggest a lobe origin for the $\gamma$ rays.
With the extended nature of the $>100$\,MeV  $\gamma$-ray emission
established, we model the source broadband emission considering
currently available total lobe radio and millimeter flux
measurements, as well as X-ray detections 
attributed to inverse Compton (IC) emission off the cosmic microwave background (CMB).
Unlike the Centaurus~A case, we find that a leptonic model
involving IC scattering of CMB
and extragalactic background light (EBL) photons underpredicts
the $\gamma$-ray fluxes by factors of about $\sim 2 - 3$, 
depending on the EBL model adopted. An additional
$\gamma$-ray spectral component is thus required, and could 
be due to hadronic emission arising from proton-proton collisions 
of cosmic rays with thermal plasma within the radio lobes.

\end{abstract}

\keywords{galaxies: active --- galaxies: individual (Fornax~A) --- galaxies: jets --- gamma rays: galaxies --- radiation mechanisms: non-thermal}

\section{Introduction}
\label{section-intro}

\noindent The radio galaxy Fornax~A, well known for its radio lobes spanning $\sim50\arcmin$,
with a lobe-to-lobe separation of $\sim33\arcmin$ \cite[see][]{ekers83}, is
one of the closest and brightest radio galaxies, located at a distance of only
$18.6$\,Mpc \citep{mad99}. Hosted by the elliptical galaxy NGC\,1316, the radio source
contains a low-ionization nuclear emission-line region nucleus, which has been imaged to arcsecond-scale resolution
and features a flat spectrum ($\alpha=0.4$; $S_{\nu} \propto \nu^{-\alpha}$) core with dual-opposing
``s''-shaped jets that are detected out to $\sim5$\,kpc from the core
\citep{gel84}.  The radio lobes are characterized by a complex polarized
filamentary structure with no observable hotspots \citep{fom89}. 

Fornax~A was the first radio galaxy reported to emit diffuse,
non-thermal X-ray emission from within its radio lobes from observations
with ROSAT \citep{fei95} and ASCA \citep{kan95}, which were later confirmed through
dedicated observations of the east lobe with 
\textit{XMM-Newton} \citep{iso06} and the west lobe with \textit{Suzaku}
\citep{tas09}. The non-thermal X-rays have been widely
attributed to inverse Compton (IC) emission of relativistic electrons scattering on
cosmic microwave background (CMB) photons, with the same population of relativistic electrons
producing both synchrotron and IC emission \citep[e.g.,][]{har79}. 
To date, similar leptonic IC/CMB emission has been detected in X-rays from tens of 
extended lobes in radio galaxies and quasars. In general, 
such detections imply that the ratio of relativistic electron pressure to 
magnetic field pressure within the lobes is $\sim1$--$100$ \citep{cro05, kat05, iso11}. More
recently, \cite{set13} reported a detection of thermal emission from the
western lobe of Fornax~A using combined \textit{Suzaku} and \textit{XMM-Newton} data.  Thermal emission
in the lobes of a radio galaxy is typically not seen, although evidence for
this has also been reported in the giant lobes of the nearby radio galaxy
Centaurus A \citep{sta13, osu13}.

Motivated by the observed (and presumed IC/CMB) X-ray emission
from the lobes of Fornax~A, \cite{che07} predicted that the
high-energy tail of the IC/CMB would be detected by the \Fermi \
Large Area Telescope \citep[LAT;][]{atw09}
at $>100$\,MeV. 
Following this, \cite{geo08} predicted that the
lobes would also be detected in $\gamma$ rays at higher energies by the LAT due to IC upscattering of the infrared and optical
extragalactic background light (EBL) photons, analogous to the
CMB photons upscattered to X-ray energies. The
association of Fornax~A with the \Fermi-LAT second year catalog
\citep[hereafter 2FGL;][]{2fgl} source 2FGL\,J0322.4$-$3717 thus
raised an important question regarding the origin of the
$\gamma$-ray source, which at the time had
no evidence presented for significant extension. In particular, a distinction between
emission arising from the lobes and possible contamination from
the central core region could not be established from the
$\gamma$-ray data alone, although X-ray and
radio observations \citep{kim03} suggested that the contribution from the core was likely
to be minimal.

In a recent study by \cite{mck15}, the spectrum of the Fornax~A lobes was modeled 
in multiple wavelengths using both leptonic and hadronic production scenarios
without knowledge of $\gamma$-ray spatial extent or 
$\gamma$-ray contamination from the galaxy core. 
They concluded the most likely
source of $\gamma$-ray production is hadronic processes within filamentary structures
of the lobes.
Our study follows the successful $\gamma$-ray detection of the extended lobes from 
Centaurus~A \citep{cenalobes}, and LAT studies of the lobes of
NGC~6251 \citep{tak12} and Centaurus~B \citep{kat13}. Gamma-ray upper limits 
using H.E.S.S. and \Fermi-LAT observations have been used to
constrain the hadronic cosmic-ray population within the radio lobes of Hydra~A \citep{hydra}.

Fornax~A is not associated with a $\gamma$-ray source in the 
most recent, third \Fermi-LAT catalog based on four years
of LAT data \citep{3fgl, 3lac}. However, the centroid of the source 3FGL~J0322.5$-$3721
is offset by $0\fdg15$ from the core of Fornax~A. This offset is greater
than the 95\% position uncertainty of the 3FGL source.
In the following we discuss possible reasons
for this offset, detail a significant $\gamma$-ray detection of extended emission from Fornax~A 
using $6.1$ years of \Fermi-LAT data, and present 
modeling under leptonic and hadronic scenarios.
Detecting extended emission from Fornax~A with the LAT is challenging
because the 68\% containment point-spread 
function (PSF) radius is $\sim0\fdg8$ at
$1$\,GeV,
which is larger than the Fornax~A lobe-to-lobe separation. The 
LAT PSF is energy dependent going from $5\deg$ at 100\,MeV to
$0\fdg1$ at 100\,GeV with 68\% confidence\footnote{\burl{http://www.slac.stanford.edu/exp/glast/groups/canda/lat_Performance.htm}}.
\\

\section{Observations \& Analysis}
\label{section-observations}

\subsection{{\it Fermi}-LAT Observations}
\label{section-lat}

Unlike all studies mentioned in Section~\ref{section-intro}, we used
$6.1$ years (from 2008 August 4 to 2014 September 4) of Pass~8 LAT data.
Compared to previous iterations of the LAT event-level analysis, Pass~8
provides greater acceptance and a PSF \citep{sgr14}, as well as 
event type partitions according to PSF\footnote{\burl{http://fermi.gsfc.nasa.gov/ssc/data/analysis/LAT_essentials.html}}, which we used in this analysis.
All of these factors allowed for a firm detection of extension of Fornax~A.
We selected from all-sky survey data at energies from 0.1 to 300\,GeV
extracted from a region of interest (ROI) with $10\deg$ radius
centered at the
J2000.0 radio position of Fornax~A \cite[R.A.\,$=50\fdg 673$, Decl.\,$=-37\fdg 208$,][]{gel84}.
We used the ``source'' event class, recommended for individual source
analysis, a zenith angle limit of $100\deg$ to greatly reduce contamination from
the Earth limb, and a rocking angle limit of $52\deg$.
\textit{Fermi} Science Tools \texttt{v10r01p00} and
instrument response functions (IRFs) \texttt{P8R2\_SOURCE\_V6} were
used for this analysis\footnote{\burl{http://fermi.gsfc.nasa.gov/ssc/data/analysis/software}}.

To model the LAT data, we
included all sources from the 3FGL
within $10\deg$ of the radio core position of Fornax~A. The diffuse
background was modeled using preliminary versions of Galactic diffuse
and isotropic spectral templates recommended by the \Fermi-LAT
collaboration, of which the finalized versions have been released to
the public\footnote{\burl{http://fermi.gsfc.nasa.gov/ssc/data/access/lat/BackgroundModels.html}}.
Several tests were performed, and we determined that the results presented here 
with the preliminary diffuse models are compatible with those obtained with the finalized models.
We used the same spectral models as in the 3FGL catalog for all background sources, and
the normalization and spectral shape parameters of all 
point sources were left free during optimization.
For the diffuse models, only the normalization parameters
were left free.

\begin{figure*}[t]
	\includegraphics[width=\textwidth]{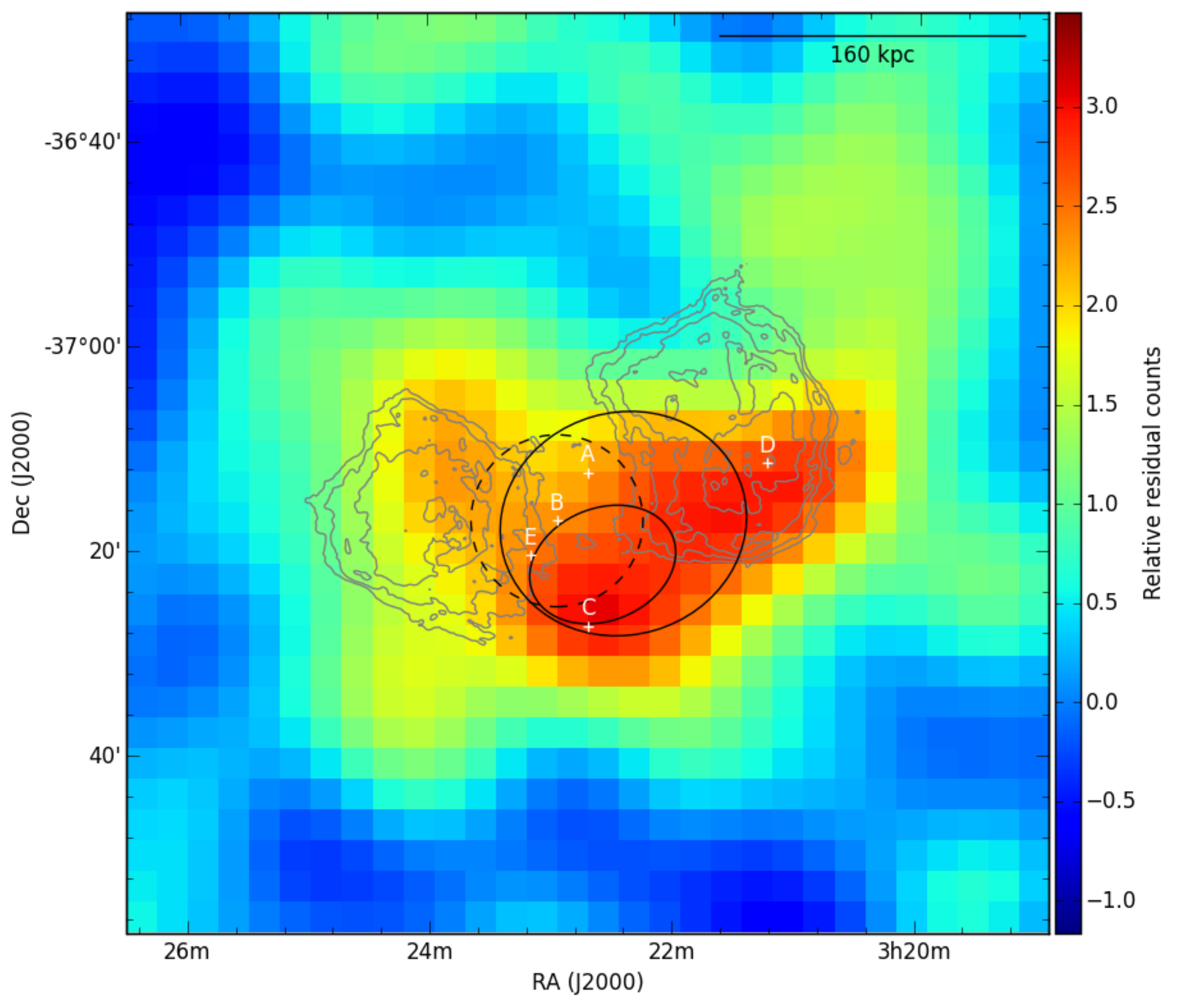}
	\caption{
		Relative smoothed (4.7 px, $0\fdg 24$ Gaussian FWHM) residual $\gamma$-ray counts
		((counts - model)/model) in the $1\fdg 5$ square region around the
		core of Fornax~A between 1 and 300\,GeV.
		Overlaid are the radio contours (gray lines) from the VLA observations of \cite{fom89} with the radio core (position indicated by A) subtracted.
		The $\gamma$-ray residual counts are elongated in a similar way to
		the radio lobes. Also shown are the 95\% confidence error ellipses for 2FGL\,J0322.4$-$3717
		and 3FGL J0322.5$-$3721; the 3FGL source has the smaller ellipse. The dashed 
		circle shows the 95\% confidence error circle from our 
		maximum-likelihood localization of the region as a single 
		point source centered at position B. The other points 
		(labeled C, D, E) are the locations of various sources and test
		sources, as detailed in the text and in Table~\ref{tab:loglike}.
	}
	\label{fig:residuals}
\end{figure*}

We initially modeled Fornax~A as a point source located at the position
of the radio core, removing 3FGL J0322.5$-$3721 from the model
since it is offset from the radio core by $0\fdg15$.
We optimized the localization using the
\texttt{gtfindsrc} tool provided in the Science Tools in unbinned
mode. The best-fit localization is R.A.\,$=50\fdg 73$, Decl.\,$=-37\fdg 28$ 
with a 95\% confidence
error circle radius of $0\fdg 14$, slightly southeast of the
position of the Fornax~A core and consistent with the reported 
3FGL localization. Figure~\ref{fig:residuals} shows the
best-fit localization (point B) and the core (point A) as well as the 2FGL and 3FGL error
contours plotted on top of the relative residual counts map.
Optimizing the model with the single point source at point B, we detect
$\gamma$-ray emission at a Test Statistic (TS)\footnote{$TS$ is defined as 
twice the difference between the logarithmic likelihood of the null 
hypothesis $\mathcal{L}_0$ and the alternative hypothesis being tested 
$\mathcal{L}_1$ \citep{mat96}: $TS = 2 (\log{\mathcal{L}_1 } - \log{\mathcal{L}_0})$.} = 121. The spectrum was modeled
as a single power law with a resulting maximum-likelihood photon index $\Gamma=2.08 \pm 0.08$ and
a full band energy flux of $(5.34 \pm
0.78_{\text{stat}} \substack{+0.03 \\ -0.05} \ _{\text{sys}} )\times10^{-12}$\,erg cm$^{-2}$ s$^{-1}$ (see Table~\ref{tab:ts}).
Systematic errors are due to the systematic uncertainty in the LAT effective area\footnote{\burl{http://fermi.gsfc.nasa.gov/ssc/data/analysis/LAT_caveats.html}}.
These fluxes and indices are consistent with those reported in the catalogs for 
sources 2FGL\,J0322.4$-$3717 and 3FGL J0322.5$-$3721.

\subsection{Extension and Morphology}
In the following, we describe several tests performed to determine
the morphology of the observed $\gamma$-ray emission from the direction of Fornax~A,
as summarized in Tables~\ref{tab:loglike}~\&~\ref{tab:ts}. All tests on extension and morphology
made full use of the additional spatial information brought about by the new Pass~8 PSF event type partitions.
The broadband flux and spectrum optimizations in addition to these tests were performed using all
PSF types in composite likelihood.

\subsubsection{Spatial Extension}
\label{section-extension}

\begin{figure}
	\centering
	\subfigure[]{
		\includegraphics[width=0.5\textwidth]{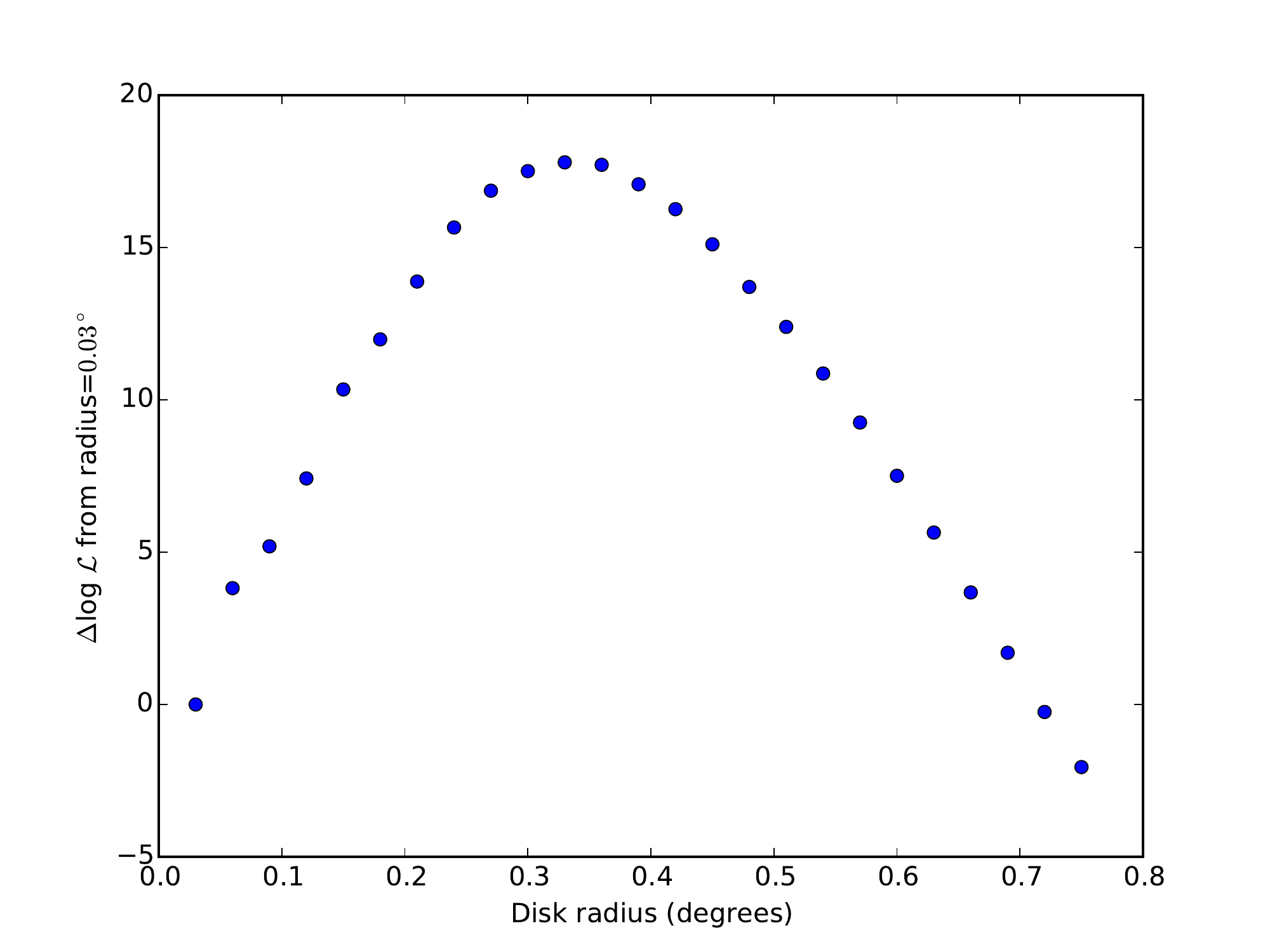}
		\label{fig:disks}
	}
	\subfigure[]{
		\includegraphics[width=0.5\textwidth]{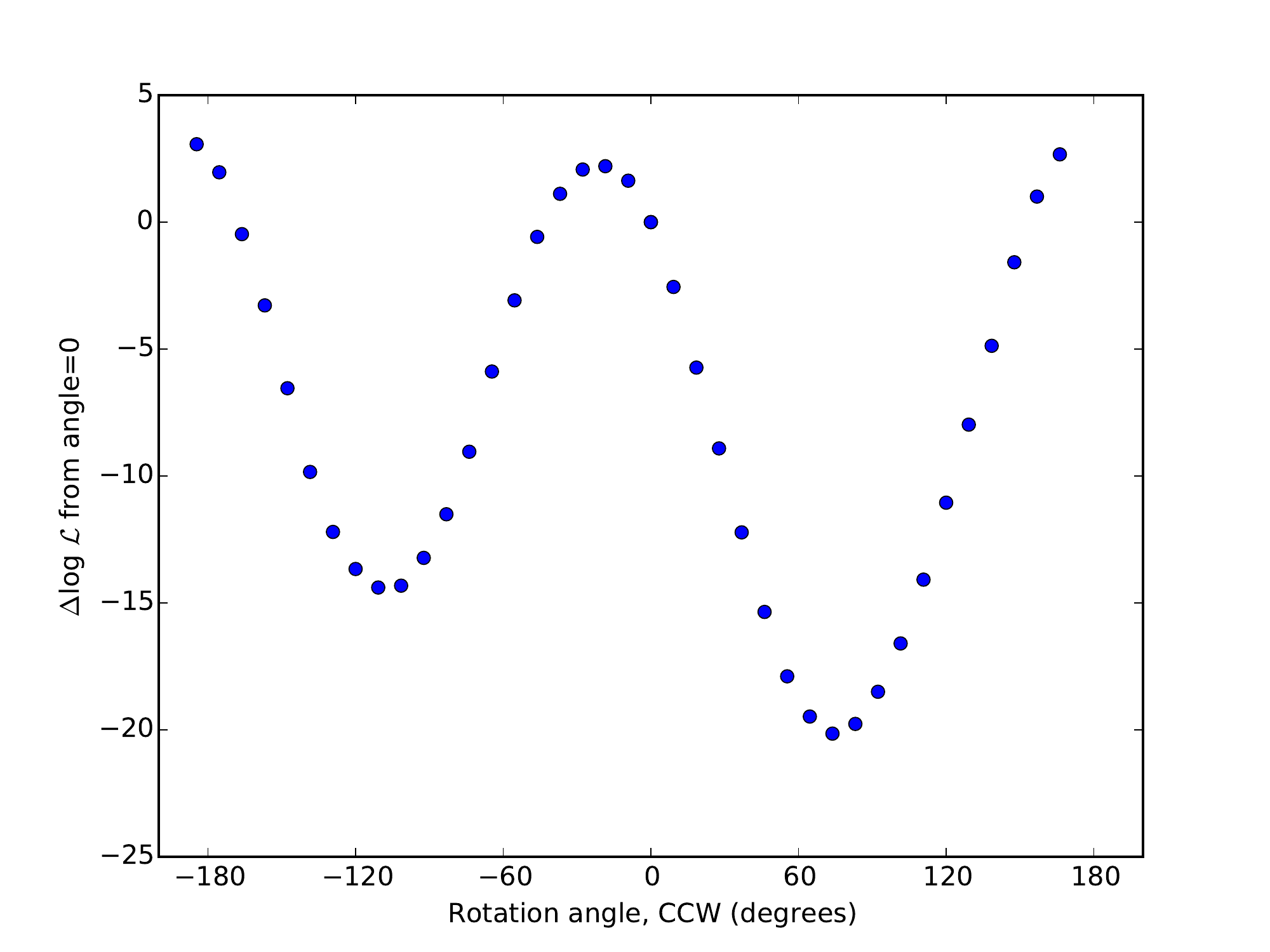}
		\label{fig:rotation}
	}
	\caption{
		(a) $\Delta \log \mathcal{L}$ between a flat disk of radius $0\fdg 03$ (i.e., point-like)
		and flat disks of various radii. A cubic fit gives a maximum likelihood radius of $0\fdg 33\pm0\fdg 05$.
		The increase in $\mathcal{L}$ from $0\fdg03$ to the maximum likelihood radius indicates the emission 
		is not point-like.
		(b) $\Delta \log \mathcal{L}$ between the non-rotated radio lobe template
		and the template rotated around the central core indicating that the unrotated radio
		morphology is preferred.
	}
\end{figure}

To determine if the $\gamma$-ray emission is extended beyond that of a point source
we modeled Fornax~A as a flat circular disk of various sizes ($0\fdg 03$ to
$0\fdg 75$ in steps of $0\fdg 03$) by producing several disk
templates centered at the best-fit location of the LAT source 
described in Section~\ref{section-lat} (point B in Figure~\ref{fig:residuals}). 
The uniform disk is the simplest spatial model, and the use of
a Gaussian profile has typically been shown to produce comparatively 
little difference in the overall likelihood and best-fit spectral parameters \citep{lan12}.
Using \texttt{gtlike} in binned mode (with bin size $0\fdg 05$), we determined the 
overall likelihood $\mathcal{L}$ as a function of the disk radius $r$.
As shown in Figure~\ref{fig:disks}, $\mathcal{L}$ is peaked at $r=0\fdg33\pm0\fdg05$,
which is roughly compatible with the extent of the lobes as observed 
in radio \cite[$1.5$\,GHz at $14\arcsec$ resolution,][]{fom89}.
By comparing the likelihood of the peak radius with the
near-zero radius of $0\fdg03$ (effectively a point source), we found that the $\gamma$-ray emission is spatial extended with $5.9\sigma$ confidence
\cite[$\Delta \log \mathcal{L}$ = 17.3, 1 degree of
freedom,][]{wilks}. See Tables~\ref{tab:loglike} and \ref{tab:ts} for more information.

\subsubsection{Blind Tests for Morphology}
\label{section-blind}

With the aim of making no prior assumptions about the morphology of
the $\gamma$-ray emission in this region, we used the ROI fit with
our position-optimized point source as described in 
Section~\ref{section-lat} and removed that point source
from the model, thereby leaving only the background sources.
Using this background model, we created
a map of residual counts above 1 GeV in the ROI by subtracting the model's
predicted counts from the observed counts in each bin. Apart from the
emission near Fornax~A, the map of residuals is flat in significance and
the distribution of significance values for the bins is Gaussian, indicating there
are no significant systematic deviations from the ROI model. 
The map of residual counts shown in Figure~\ref{fig:residuals}
is cropped to a $1\fdg 5\times1\fdg 5$ region centered around
the Fornax~A core.  The shape of the residuals suggests a non-circular
morphology that resembles the known extended radio morphology.
The contours of radio emission from a VLA $1.5$\,GHz image at
$14\arcsec$ resolution \citep{fom89}, with the radio core subtracted, are
overlaid in Figure~\ref{fig:residuals}.

We also tested a model which included two separate point sources whose initial
locations were chosen by eye based upon the residual counts. The locations 
of these two point sources were then optimized using
\texttt{gtfindsrc}.  The best-fit location of the western point source (labeled 
D in Figure~\ref{fig:residuals}) matches well the western lobe's 
centroid while the eastern point (labeled E in Figure~\ref{fig:residuals}) 
is offset from the centroid of the eastern lobe.
The double point-source model is preferred over the single point-source model at a
confidence level of $4.8\sigma$. Table \ref{tab:loglike} details these
results and Table \ref{tab:ts} lists the fit parameters.  Note that
the spectral index is statistically compatible with the single point-source
model and for both point sources in the double point-source model.

\subsubsection{Radio-Motivated Tests for Morphology}
\label{section-radio}

Assuming that the same electron energy distribution determined from the radio
emission scatters optical EBL photons to produce $\gamma$ rays, then a reasonable guess 
for the $\gamma$-ray morphology should be the observed radio structure.
The lobes of Fornax~A were found to emit
non-thermal IC scattered X-rays, with excellent spatial coincidence
to the radio structure produced by synchrotron-emitting relativistic electrons \citep{fei95}.
Diffuse hard X-ray emission associated with the east lobe has been
confirmed with \textit{XMM-Newton} \citep{iso06}, implying the magnetic field
is reasonably uniform, further supporting the expectation that the $\gamma$ rays should
match the radio morphology. Under these circumstances, as was observed in
Centaurus A \citep{cenalobes}, the relativistic electrons will trace the
$\gamma$-ray emission \citep{geo08}. These assumptions could be incorrect;
however, for the purpose of constraining the EBL with the $\gamma$-ray flux,
using the radio structure is the best choice for the spatial distribution.
Physically, Fornax~A images from the Wilkinson Microwave Anisotropy 
Probe (\textit{WMAP}) might be a better choice of template for the $\gamma$-ray
emission because the synchrotron radiation within $\sim23$\,--\,$94$\,GHz should
be originating from the same band of relativistic electrons that IC scatter into
$\sim0.1$\,--\,$10$\,GeV $\gamma$ rays. We revisit this topic later in the section.

We created a spatial template of the lobe emission using VLA $1.5$\,GHz radio
data \citep{fom89}, which offers a more-than-adequate spatial resolution of $14\arcsec$. Both the central core of Fornax~A
and various radio point sources inside the lobe structure were manually removed
from the spatial template, the former by reduction to zero and the latter 
by interpolation of the adjacent lobe-dominated flux. The central core was reduced to zero
because the lobes do not overlap with the central core \citep{ekers83}.
We found this template in addition to
a point-source model of the core is preferred over just the point source
at the core with a confidence of $6.0\sigma$ ($\Delta \log \mathcal{L}$ = 19.8, 2 degrees of
freedom). However, this combined
model is preferred over the lobes template alone by only $0.7\sigma$.
Moreover, in the combined
point-source-and-lobes template fit only 14\% of the energy flux is assigned
to the core point source.
See Tables~\ref{tab:loglike} and \ref{tab:ts} for more information.
We consider this as evidence that the $\gamma$-ray emission from
the core of Fornax~A is insignificant. Minimal $\gamma$-ray flux
from the central core is expected, as it is assumed to be synchrotron self-Compton,
and the radio flux from the core has been reported to be relatively weak compared
to the luminous galaxy lobes \citep{gel84}.

In order to test the uniqueness of the radio template morphology and
its rotational symmetry, we rotated the template around the central core in
increments of $9\fdg2$ and computed the likelihood at each one. The results from this test
are shown in Figure~\ref{fig:rotation} and indicate that
the template in its original orientation is preferred. 
In particular, we see the original orientation is a
better model than the one rotated $90^{\circ}$ clockwise 
with $5.1\sigma$ confidence ($\Delta \log \mathcal{L}$=12.9, 1 degree of freedom)
and $90^{\circ}$ counterclockwise with
$6.1\sigma$ confidence ($\Delta \log \mathcal{L}$=18.9, 1
degree of freedom). The maximum likelihood
rotation was $-20^{\circ} \pm 10\fdg$ Additionally, our plot of $\log \mathcal{L}$ over
template rotation angle shows a sinusoidal profile with a similar peak
in likelihood around $180^{\circ}$ as around $0^{\circ}$.
This degeneracy indicates that the data are not constraining enough to statistically
differentiate the lobes. Modeling the lobes as separate point sources (see Section~\ref{section-blind})
results in the measurement of similar fluxes and spectral shapes for the two regions,
but this is at odds with radio observations of the lobes.
According to the $1.5$\,GHz VLA observation \citep{fom89}, the west lobe has about 
twice the total flux than the east. In fact, the two point-source $\gamma$-ray analysis indicated
the east lobe may be slightly brighter (but is within the statistical errors).
This disagreement may be a symptom of our use of $1.5$\,GHz VLA data instead of
the very similar but more physically motivated $\sim23$\,--\,$94$\,GHz \textit{WMAP} data
as a spatial template for the $\gamma$-rays. Indeed, the flux ratio of west to east
is $\sim1.3$ in the \textit{WMAP} $41$ and $61$\,GHz maps \citep{geo08}, closer to the
$\gamma$-ray result. However, Fornax~A is at the resolving power of the LAT in this analysis
and the PSF is broad enough ($\sim0\fdg8$ radius at $1$\,GeV) that fine scale changes would be smoothed out
and indistinguishable from the original, and small changes in the flux ratio between the lobes should
leave the average flux nearly the same. This is highlighted by the 
similarity in overall likelihood we observe between $0^{\circ}$ and $180^{\circ}$ rotations
of the $1.5$\,GHz VLA template, where the rotation of $180^{\circ}$ is identical to 
a flux ratio of $\sim0.5$. In addition, even when using the spatial morphology
of a single point source, the flux is nearly consistent 
with the radio template flux (see Section~\ref{section-lat}).

With the aim of testing the region for the possibility of a contaminating
background $\gamma$-ray source, we added a point source to the lobes template model
and optimized its position using \texttt{gtfindsrc}. The optimized position
of this point source (labeled C in Figure~\ref{fig:residuals}) was 
R.A.~=~$50\fdg67$, Decl.~=~$-37\fdg46$ with a 95\% confidence 
error circle radius of $0\fdg42$ (large enough to encompass the 
whole Fornax~A emission region). The resulting fit was marginally
preferred over the lobes template alone with $2.7\sigma$ significance.
Therefore, we do not consider any contribution from a background point source to be significant.

\begin{deluxetable*}{llrrrr}
	\tablecolumns{5}
	\tablewidth{0pc}
	\tablecaption{Fornax~A spatial model comparisons\label{tab:loglike}}
	\tablehead{
		\colhead{Null hypothesis} &
		\colhead{Alternative hypothesis} &
		\colhead{DOF\tablenotemark{a}} &
		\colhead{$\Delta \log \mathcal{L}$} &
		\colhead{$\sigma$\tablenotemark{b}} &
		\colhead{Sect.}}
	\startdata
		Point source (core location A) & Point source (best-fit location B) & 2 & 0.6 & 0.6 & \ref{section-lat}\\
		Disk (best-fit location B, $0\fdg 03$ radius) & Disk (best-fit location B, $0\fdg 33$ radius) & 1 & 17.3 & 5.9 & \ref{section-extension} \\
		Point source (best-fit location B) & Two point sources (locations D and E) & 4 & 16.2 & 4.8 & \ref{section-blind} \\
		Point source (core location A) & Radio lobes template and point source (core location A) & 2 & 19.8 & 6.0 & \ref{section-radio} \\
		Radio lobes template & Radio lobes template and point source (core location A) & 2 & 0.7 & 0.7 & \ref{section-radio} \\
		Radio lobes template, rotated $90^{\circ}$ CW & Radio lobes template & 1 & 12.9 & 5.1 & \ref{section-radio} \\
		Radio lobes template, rotated $90^{\circ}$ CCW & Radio lobes template & 1 & 18.9 & 6.1 & \ref{section-radio} \\
		Radio lobes template & Radio lobes template and point source (location C) & 4 & 7.0 & 2.7 & \ref{section-radio} \\
	\enddata
	\tablecomments{Calculated using the likelihood ratio computation as described in Section~\ref{section-observations}.}
	\tablenotetext{a}{The difference in the number of degrees of freedom between the two hypotheses.}
	\tablenotetext{b}{The alternative hypotheses are preferred over the corresponding null hypotheses by the significances $\sigma$.}
\end{deluxetable*}

\begin{deluxetable*}{lrcc}
	\tablecolumns{4}
	\tablewidth{0pc}
	\tablecaption{Fornax~A LAT spectral fit results\label{tab:ts}}
	\tablehead{
		\colhead{Model} &
		\colhead{TS} &
		\colhead{Energy flux ($\times10^{-12}$\,erg cm$^{-2}$ s$^{-1}$)} &
		\colhead{Photon index}}
	\startdata
		Point (best-fit location B) & 121 & $5.34 \pm 0.78 \substack{+0.03 \\ -0.05} \ _{\text{sys}}$ & $2.08 \pm 0.08 \pm 0.03_{\text{sys}}$ \\
		Disk (best-fit location B, $0\fdg 33$ radius) & 158 & $7 \pm 1$ & $1.99 \pm 0.07$ \\
		Radio lobes template & 158 & $7.57 \pm 1.05 \substack{+0.06 \\ -0.08} \ _{\text{sys}}$ & $1.99 \pm 0.07 \substack{+0.03 \\ -0.04} \ _{\text{sys}}$ \\
		Two point sources, summed & & $6.6 \pm 0.8$ & \\
		\hspace{1cm} West (location D) & 37 & $2.9 \pm 0.7$ & $2.02\pm0.13$ \\
		\hspace{1cm} East (location E) & 51 & $3.7 \pm 0.8$ & $2.05\pm0.11$ \\
		Radio lobes template and point source (location C), summed & & $8 \pm 1$ & \\
		\hspace{1cm} Radio lobes template & 66 & $6 \pm 1$ & $1.97 \pm 0.08$ \\
		\hspace{1cm} Point source (location C) & 14 & $1.6 \pm 0.6$ & $2.01 \pm 0.20$ \\
	\enddata
	\tablecomments{Each fit was performed leaving normalization and spectral shape parameters of all sources free, except for the models for the diffuse background $\gamma$ rays which were fit with only normalization.}
\end{deluxetable*}

\subsection{Spectral and Temporal Analysis}
\label{subsec:sed+lc}

In the following we assume the radio morphology template (without any core
contribution) is the best description of the Fornax~A $\gamma$-ray
emission. The likelihood ratio technique cannot quantify whether
the radio template is statistically preferred with respect to 
the best-fit disk model because these models are not nested.
However, the radio template is the physically motivated model based on
the leptonic scenario of $\gamma$-ray production. Further, the rotation
study presented above indicates a preferred axis of the $\gamma$-ray emission
which mimics the elongated emission observed at radio frequencies, in turn
supporting the similarity between the $\gamma$-ray and radio emission morphology.

\begin{figure*}[t]
	\includegraphics[width=\textwidth]{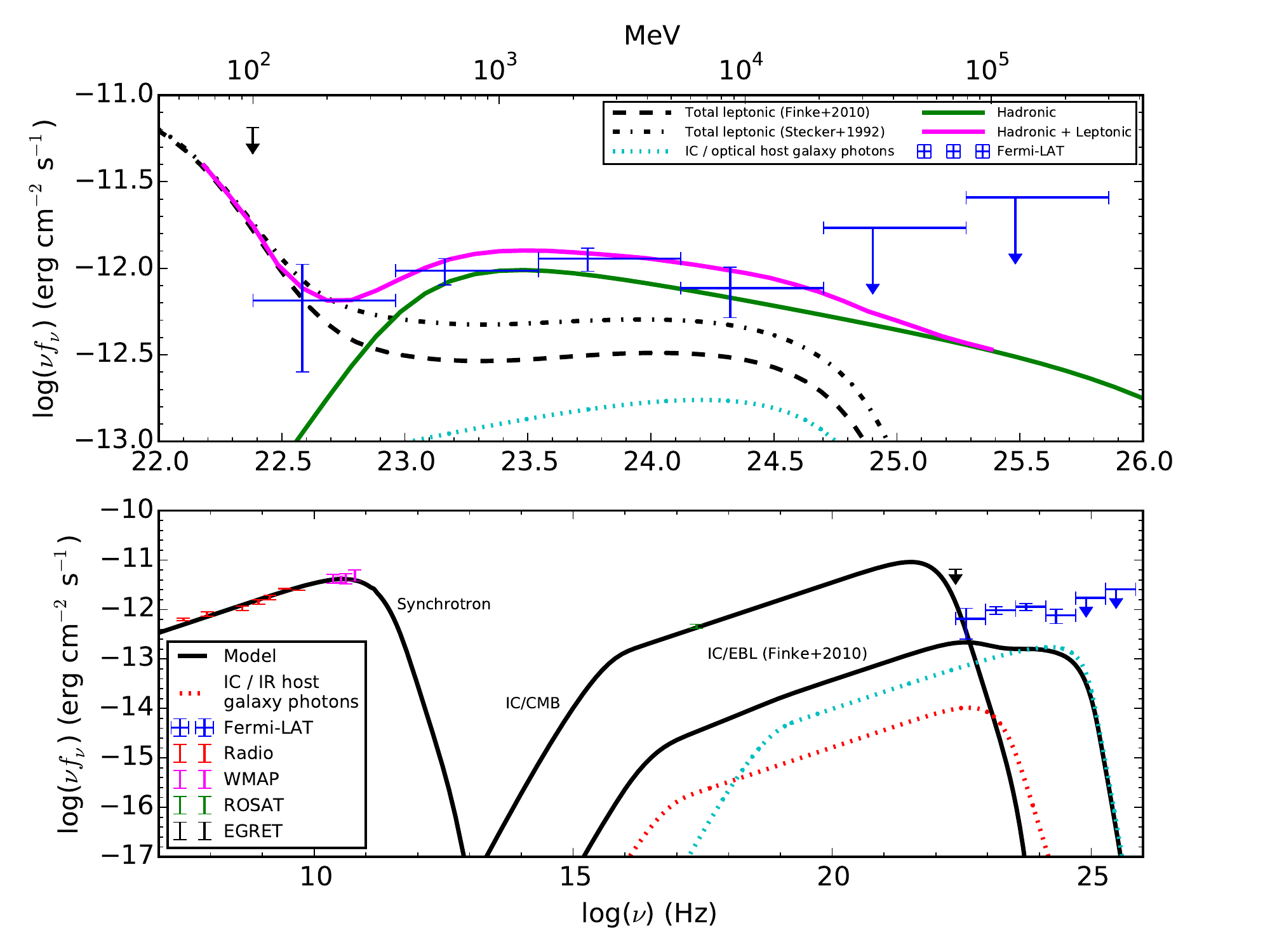}
	\caption{
		Broadband SED of the lobes of Fornax~A (bottom panel) and detailed
		view of the high-energy part of the SED (top panel). As in \cite{geo08},
		we used archival measurements of the total lobe radio flux densities (shown in red)
		from \cite{fin73}, \cite{cam71}, \cite{jon92}, \cite{ekers83}, \cite{bol73}, and \cite{kuh81},
		as collected by \cite{iso06}, replacing an extrapolated 100\,MHz data point from \cite{fin73}
		with an 86\,MHz measurement \citep{mil60}.
		The 3-year integrated \textit{WMAP} data are shown in magenta \citep{hin07}, 
		and X-ray data from ROSAT \citep{fei95} are shown
		in green. The LAT data points from this study are shown in blue. The black upper limit point is from EGRET \citep{cil04}.
		The black solid lines from left to right show the synchrotron ($<10^{13}$\,Hz), IC/CMB ($>10^{13}$\,Hz), and
		IC/EBL \citep[on the assumption of the model of][]{fin10} flux models ($>10^{15}$\,Hz). The dashed black line 
		shows the IC/EBL flux models assuming the fast evolution model of \cite{stecker}.
		The dotted red and cyan lines show the IC upscattered 
		host galaxy photon flux from infrared and optical, respectively.
		The solid green line shows the hadronic model flux, while the solid magenta line shows
		the combined hadronic and leptonic model flux.
		}
	\label{fig:sed}
\end{figure*}

We tested several broadband spectral models (log-parabola, broken power law,
and broken power law with an exponential cutoff), and found none were significantly preferred over
the single power law. Then, we measured spectral points by fitting each of 6 equal logarithmically
spaced energy bins from 0.1 to 300\,GeV to a power law and optimized the flux normalizations by maximizing the likelihood function.
In each bin, the normalization parameters for all sources were free,
and all other parameters were fixed to the values obtained from the broadband fit.
These spectral data points are shown in Figure~\ref{fig:sed} and Table~\ref{tab:sed}.
The source is detected in 4 of the 6 spectral bins with TS $>5$,
and 95\% confidence upper limits were calculated for the 
two lower-significance bins (at the highest energies).

\begin{deluxetable*}{ccc}
	\tablecolumns{3}
	\tablewidth{0pc}
	\tablecaption{Fornax~A total lobe LAT spectral flux\label{tab:sed}}
	\tablehead{
		\colhead{Bin energy range (GeV)} &
		\colhead{Energy flux ($\times10^{-12}$\,erg cm$^{-2}$ s$^{-1}$)} &
		\colhead{TS}}
	\startdata
		$0.10$ -- $0.38$ & $0.7 \pm 0.4$ & \phn6.1 \\
		$0.38$ -- $1.4$\phn & $1.0 \pm 0.2$ & 49\phd\phn \\
		$1.4$ -- $5.5$ & $1.1 \pm 0.2$ & 87\phd\phn \\
		$5.5$ -- $21$\phd & $0.8 \pm 0.2$ & 22\phd\phn \\
		\phm{*}$21$ -- $79^*$ & $<1.7$ & \phn3.5 \\
		\phm{*}\phn$79$ -- $300^*$ & $<2.6$ & \phn4.3 \\
	\enddata
	\tablecomments{Assuming radio lobes template spatial model}
	\tablenotetext{*}{95\% confidence upper limits}
\end{deluxetable*}

To test the $\gamma$-ray variability over the $6.1$ year period,
we made a $0.1$--$300$\,GeV light
curve in time bins of 185 days, which was found to be the smallest
possible time scale while maintaining a reasonable significance of detection in
the majority of bins. For each time interval, the emission associated with Fornax~A was fit to a single point source (positioned at the best-fit location B in Figure~\ref{fig:residuals}),
as we expected any potential variable emission to be 
associated with a point source at the core and not the lobes.
All sources included in the $6.1$ year analysis were
fit with all spectral shape parameters fixed to their optimized values from the full fit, 
while all normalizations were left free.
Upper limits were calculated for time bins within which the TS fell below 4 ($<2 \sigma$). The significance of variability
was determined following the method described in \cite{2fgl}. Our analysis yielded a 
$1.3\sigma$ confidence that the emission is variable, and so we conclude that
we do not observe significant variability.

\section{Results \& Discussion}
\label{section-discussion}

Our \Fermi-LAT study of the region around Fornax~A consistently shows that,
under all tests performed, the $\gamma$-ray emission region is significantly extended and the most likely spatial distribution is
delineated by the radio lobes. Using a size-optimized flat disk model,
extension beyond a point source was found to be significant at $5.9\sigma$
confidence, with a preferred
radius of $0\fdg 33 \pm 0\fdg 05$. Modeling the emission as
two point sources results in a western point source well matched to the radio
lobe centroid and an eastern point source offset from the eastern lobe.  This
model is preferred over a single point source at the $4.8\sigma$
level. Furthermore, using the $1.5$\,GHz VLA radio morphology \citep{fom89} as a template in combination
with a central core point source results in a significantly greater likelihood than the point source alone
with $6.0\sigma$ confidence.
Contamination from the core is determined to be at most
14\% based on a likelihood fit with the radio lobes template and 
a point source at the core location. While it is difficult to determine
the exact morphology of the $\gamma$-ray emission,
our study shows that it cannot be fully described as a point-like source.

A few scenarios could explain the offset $\gamma$-ray point-source localization
seen in the 3FGL and in the single point-source analysis presented
here. Firstly, since we now know the emission is extended (or at least not point-like),
to use a point-source model to localize the emission is to start with a false assumption.
The distribution of the $\gamma$-ray emitting regions
may not be uniform across the lobe structure and thus would
not result in a symmetric distribution of $\gamma$-ray emission. 
Second, based upon the offset
eastern lobe point-source localization (point E in Figure~\ref{fig:residuals}), the existence of a background
$\gamma$-ray source is not ruled out. However, adding a point
source to the lobes template model and localizing with \texttt{gtfindsrc} 
yields only a slightly better fit at
the $2.7\sigma$ level (see point C in Figure~\ref{fig:residuals}).
We also find no evidence for variability in this source over $\sim6$ years of observations.
Variability might support the presence of a common background source such 
as a blazar. We note that another potential source that has been investigated in this region is the Fornax cluster \citep{and12,ack14} whose center 
lies $3\fdg6$ northeast of the Fornax~A core, and may be contributing
contaminating $\gamma$-ray flux from various cluster constituents.
However, no galaxy cluster has been detected in $\gamma$ rays so far.

\Fermi-LAT data have been previously used by \cite{mck15} to study Fornax~A.
They reported a photon flux above $100$\,MeV of $6.7\times10^{-9}$\,ph~cm$^{-2}$~s$^{-1}$ using a point
source spatial model, and our finding using the lobes template was close at $(5.7\pm0.9)\times10^{-9}$\,ph~cm$^{-2}$~s$^{-1}$.
Our study establishes for the first time spatial extension of Fornax~A in $\gamma$ rays,
and distinguishes between $\gamma$-ray contributions from
the core and lobes.
This result was enabled thanks to 
the improvements brought about by the new Pass~8 event reconstruction, rather than the marginal
increase in exposure time (i.e., 6 years of data in our study instead of 5 years in theirs).

\subsection{Leptonic Modeling}
We model the $\gamma$-ray emission following \cite{geo08}, in which
the relativistic electrons in the lobes of the radio galaxy are
IC scattered off of CMB and EBL photons.  For a given electron energy
distribution (EED) of the lobes, the resulting IC emission
will consist of a lower-energy component due to CMB photons, as well
as two components at higher energy due to the cosmic infrared and 
optical backgrounds (CIB and COB, respectively). While the
EBL energy density is only a few percent that of the CMB, 
the resulting IC spectrum due to CIB and COB photons will be shifted in frequency
by $\gamma^{2}_{\rm max}$, where $\gamma_{\rm max}$ is the maximum
Lorentz factor of the EED. The EED used in this model is a broken 
power law that breaks at $\gamma_{break}=1.3\times10^5$ 
from an electron index of $2.3$ to a much larger 
value to mimic a cutoff. This break was chosen 
so as to not overproduce the emission in the lowest-energy LAT band.
With peak wavelengths of $\lambda \sim 100 \
\upmu$m and $\sim 1 \ \upmu$m for the CIB and COB, the resulting IC spectrum
will be shifted in frequency by factors of $\sim 10$ and $1000$,
respectively from that of the upscattered CMB. 

This model is shown along with the $\gamma$-ray spectral energy
distribution and the radio-to-sub-mm measurements 
of the total emission from the lobes presented in \cite{geo08}
in Figure \ref{fig:sed}.
Note that because Fornax~A is an extended source
in other wavelengths as well, care must be taken in defining the spatial structure
in all wavelengths in order to draw meaningful comparisons.
This model makes use of currently available \textit{total} lobe fluxes.
\textit{WMAP} and Planck fluxes reported by \cite{mck15} were 
obtained using resolution-dependent apertures that did 
not fully enclose the extent of the synchrotron lobe emission.
We assumed a lobe magnetic field strength of 1.65\,$\upmu$G, 
constrained to the X-ray flux data point, and we include photon contributions
from the host galaxy following \cite{geo08}. 
The extracted spectral data points do not
appear to match the predicted model shape based on IC/EBL emission alone.
Fully accounting for the \Fermi-LAT observed fluxes under the
IC/EBL hypothesis alone would imply an EBL level that is even higher than the Stecker
model \citep{stecker}, which was ruled out by \cite{fermiebl}.
Consequently, the applied leptonic model cannot completely
explain the observed emission. The model relies upon the assumption that
all of the X-ray flux observed from the lobes is created by IC/CMB
scattering to obtain the magnetic field strength. If some amount of the X-ray flux is
thermal emission \cite[as reported by][]{set13}, our expected IC/EBL level would decrease, creating
further discrepancy between model and data.

As discussed in Section~\ref{section-radio}, a more physically motivated choice
of spatial template would be the higher-frequency \textit{WMAP} data. However, the 
resulting changes in flux and spectral shape should be within the statistical errors 
of our current results, and therefore should not alter our result that
the flux exceeds the leptonic model of the Fornax~A lobes.

Note that the intensity of the IC contribution from the host galaxy photons 
of Fornax~A is comparable to that of the EBL photons in the lobes, 
and it actually dominates at higher energies ($>1$\,GeV, see 
Figure~\ref{fig:sed}). This differs from the case of Centaurus~A,
wherein the predicted EBL photon intensity is roughly five times 
that of the starlight \citep{cenalobes}. Were it true that host 
galaxy photons dominate in Fornax~A, the expected spatial 
distribution of $\gamma$ rays from the lobes would not be 
uniform, with brighter emission nearer the center and less 
away from the core. Testing for this feature requires spatial 
resolution that is beyond the capabilities of the LAT with current statistics.

\subsection{Hadronic Modeling}
The problem of the model not fitting the $\gamma$-ray spectrum in Fornax~A 
may be solved by an additional contribution from hadronic cosmic rays interacting within 
the lobes, as found by \cite{mck15}. We created a model of hadronic emission (proton-proton interactions) 
assuming a total emitting volume of $7 \times 10^{70}$\,cm$^3$ 
and a uniform distribution of thermal gas with
number density $3 \times 10^{-4}$\,cm$^{-3}$ following \cite{set13}, and 
a power-law cosmic-ray spectrum with energy index~2.3 extending from $\sim3$\,GeV up to
more than 10\,TeV. Modeling the $\gamma$ rays as entirely hadronic in origin
requires a large total cosmic-ray energy of $\sim 1 \times 10^{61}$\,erg, which is
twice the observed energy of $\sim5\times 10^{60}$\,erg in the lobes of 
comparable radio galaxy Hydra A \citep{hydra} and
very high compared to  
an estimate of $\sim5\times10^{58}$\,erg in the outburst that is assumed 
to have created the lobes of Fornax~A \citep{lan10}.
We then subtracted the lowest IC/EBL model \citep{fin10} from our LAT spectral
points and fit the residual flux as hadronic emission, 
and found we could achieve a reasonable fit, 
shown in Figure \ref{fig:sed}.
The resulting cosmic-ray pressure fitted from this residual flux is $\sim 2 \times 10^{-11}$\,dyn\,cm$^{-2}$ and the
total energy stored in cosmic rays is $\sim 5 \times 10^{60}$\,erg. This total energy 
is similar to that of Hydra A \citep{hydra} and closer to an estimate of the 
total energy in the Fornax~A lobes \citep{lan10}. This result agrees with 
analogous calculations by \cite{mck15}, in which the discrepancy is explained
by suggesting the emission is primarily hadronic and localized to 
relatively denser sub-structures within the lobes, thereby 
decreasing the effective emitting volume. 

\section{Conclusions}
\label{section-conclusions}

We report the first \Fermi-LAT detection of
extended $\gamma$-ray emission from the radio galaxy
Fornax~A using $6.1$ years of \Fermi-LAT data.
We conclude that a point-source spatial model is insufficient to describe
the $\gamma$-ray emission, and our analysis indicates it is likely
the emission originates in the lobes.
We investigated the
origin of the extended emission by assuming leptonic emission
that arises due to IC scattering of EBL photons 
off of relativistic electrons in the radio lobes.
This leptonic modeling underestimates the observed $\gamma$-ray 
emission for any current EBL estimate, consistent with the recent findings by
\cite{mck15}, even after accounting for the additional contribution of IC emission
off of the host galaxy light.
A hadronic-only model (proton-proton interactions)
requires implausibly large total cosmic-ray energy when compared to an 
estimate of the Fornax~A outburst assumed to have created the lobes \citep{lan10},
and this problem can be alleviated by invoking denser sub-structures in the lobes \citep{mck15}.
When we assume the lowest EBL model \citep{fin10} and
fit the residual $\gamma$-ray flux as hadronic production, our fit yields a total cosmic-ray energy of
$\sim 5 \times 10^{60}$\,erg, matching well with the Hydra A energy \citep{hydra}, but still
at least 100 times greater than the estimated total energy in the Fornax~A lobes \citep{lan10}.
Thus, even the combined leptonic and hadronic scenario may not be able to
explain the $\gamma$-ray lobe emission.
Given our current understanding of the content of the radio lobes and the EBL, 
some contribution from leptonic processes must exist. If it is true
that there is a $\gamma$-ray component other than leptonic in Fornax~A,
we should expect to observe such a component in other nearby radio galaxies as well.

Our spatial analysis hinted at the existence of a background $\gamma$-ray source
(C in Figure \ref{fig:residuals}). We re-evaluated the SED of Fornax~A including this source in the model, but we found this background source addition is 
only marginally preferred and cannot fully 
make up the difference between the data and the IC/EBL model.
In any case, $\gamma$-ray contamination from an unresolved 
background source or the Fornax cluster could be present.

Our modeling was done using previously published multiwavelength data.
Further analysis in other wavelengths in the future will yield a more definitive picture
of the SED. More detailed observations, such as with the
hard X-ray telescope \textit{NuSTAR}, would help determine if the
X-ray emission is contaminated by thermal processes \citep{set13}, which
would test our initial assumptions that the radio data traces the
X-rays and $\gamma$ rays, and that the X-rays could be used to constrain the 
lobe magnetic field. \Fermi-LAT analysis at even lower energies (below
100 MeV) may provide more information about the IC/CMB component \citep{che07}.
Potentially with greater statistics, the \Fermi-LAT
could extend the lobes' detection to higher energies and additionally
observe the effect of the host-galaxy photons on the spatial distribution
of $\gamma$ rays in the Fornax~A lobes.

\acknowledgments

The \textit{Fermi} LAT Collaboration acknowledges generous ongoing support
from a number of agencies and institutes that have supported both the
development and the operation of the LAT as well as scientific data analysis.
These include the National Aeronautics and Space Administration and the
Department of Energy in the United States, the Commissariat \`a l'Energie Atomique
and the Centre National de la Recherche Scientifique / Institut National de Physique
Nucl\'eaire et de Physique des Particules in France, the Agenzia Spaziale Italiana
and the Istituto Nazionale di Fisica Nucleare in Italy, the Ministry of Education,
Culture, Sports, Science and Technology (MEXT), High Energy Accelerator Research
Organization (KEK) and Japan Aerospace Exploration Agency (JAXA) in Japan, and
the K.~A.~Wallenberg Foundation, the Swedish Research Council and the
Swedish National Space Board in Sweden.
 
Additional support for science analysis during the operations phase is gratefully acknowledged from the Istituto Nazionale di Astrofisica in Italy and the Centre National d'\'Etudes Spatiales in France.

{\it Facility:} \facility{\Fermi}

{}

\clearpage

\end{document}